\documentclass[preprint]{elsarticle}

\usepackage{lineno,hyperref}
\modulolinenumbers[5]

\journal{Journal of Materials Science}

\begin{document}

\begin{frontmatter}

\title{Reactive infiltration: identifying the role of chemical reactions, capillarity, viscosity and gravity}

\author{E. Louis$^{1,2}$, J.A. Miralles$^1$, J.M. Molina$^{1,2,3}$}

\address{
$^1$Departamento  de  F\'{\i}sica Aplicada,
  Universidad de Alicante, Ap. Correos 99, 03690 Alicante,
  Spain\\
$^2$Instituto Universitario
de Materiales de Alicante (IUMA) and Unidad
  Asociada del Consejo Superior de Investigaciones Cient\'{\i}ficas,
  Universidad de Alicante, Apdo. Correos 99, 03690 Alicante,
  Spain\\
  $3$ Departamento de Qu\'{i}mica Inorg\'{a}nica de la Universidad de Alicante, Universidad de Alicante, Apdo. Correos 99, E-03080 Alicante, Spain.}
 
\begin{abstract}
A wealth of experimental data indicate that while capillarity controlled infiltration gives an infiltration length that varies with 
the square root of time, reactive infiltration is characterised by a linear relationship between the two magnitudes. In addition the infiltration rate in the latter is at least two orders of magnitude greater than in the former.
 This work is addressed to investigate imbibition of a non-wetting, albeit reactive,  liquid into a capillary,  within the framework of a simple model that includes capillarity effects, viscosity and gravity. The capillary  radius is allowed to vary, due to reaction, with both position and time, according to either  an  interface  or  a diffusion law. The model allows for capillary closure when reaction kinetics dominates imbibition. At short times, and depending on whether infiltration is capillarity or gravity controlled, the infiltrated length varies  either as the square root or linearly with time. This suggest the following track for reactive infiltration:  i)  In most cases, the contact angle is initially larger than  $90^\circ$, ii) after some time,  reaction gradually replaces the interface liquid/preform by the liquid/reaction product interface and, concomitantly, the contact angle  gets closer to $90^\circ$, iii) beyond that time, gravity triggers infiltration  (actually  the contact angle does not need to be smaller than $90^\circ$ for the initiation of infiltration due to the metallostatic pressure exerted by the liquid metal on top of the porous preform), iv) thereafter infiltration is controlled by viscosity and gravity, provided that, due to reaction, the contact angle remains close to that at which infiltration was initiated.

\end{abstract}

\end{frontmatter}

\linenumbers



\section{Introduction}

Reactive infiltration is rapidly becoming a powerful method for fabrication of composites or monolithic materials with a tailored microstructure and/or final-use shape \cite{WA75,ST99,As98,As02,AS05,BV06,VB08}. A full description of reactive infiltration requires including, besides the standard ingredients of imbibition capillarity (viscosity and wetting), gravity, temperature effects, variable contact angle (depending, for instance on infiltration velocity) and, mainly as a consequence of reaction, non-wetting/wetting transition and a pore, or capillary, whose shape and size  vary with  time and infiltrated length. The simplest model describing imbibition capillarity was proposed during the first two decades of the past century \cite{BC06,Lu18,Wa21}. The model considers a cylindrical capillary in which capillarity forces the entrance of a liquid at a rate controlled by  viscosity,  predicting an infiltrated height  proportional to the square root of time. Although this law is commonly known as Washburn's law, it should be pointed out here that, as noted by Reyssat et al \cite{RC08} and more recently by Gorce et al \cite{GH16}, this law should be referred to as the BCLW imbibition law as it was discovered by other authors \cite{BC06,Lu18} well before  Washburn \cite{Wa21}.

In spite of the great importance of all types of capillary imbibition in a wide range of research and industrial areas,  work carried out on the particular system relevant to reactive infiltration is rather scarce, in comparison to  pressure assisted infiltration, see Refs. \cite{LW15,LM15,GL99}. While in the last ten years research has allowed to understand  rather satisfactorily  imbibition into cavities of arbitrary shape \cite{RC08,GH16} and, in addition,  imbibition into porous media or capillaries with voids of size varying with time has also been studied \cite{As02}, the case most relevant to reactive infiltration, namely,  systems with voids of size varying with both time and position, has only been very recently studied by means of a numerical method, the so-called Lattice-Boltzmann (LB) approach \cite{Ch08,SG15,SC16}. 

In this work imbibition of a liquid into a capillary whose radius varies with height and time is investigated including, besides capillarity effects, viscosity and gravity. 
The capillary radius is assumed to vary according to one of the two most widely used laws: an interface-like law characterised by a linear relation between radius and time, and a diffusion-like law that assumes a square root relationship between those two magnitudes \cite{AS05}. Each of these laws has its own reaction constant that may strongly depend on the actual system. Numerical results are presented for system parameters varied around those corresponding to liquid Si imbibition into carbon capillaries. Two limiting cases are considered:  capillarity control and gravity control imbibition. While the former leads to very high imbibition rates, the latter show rates several order of magnitude lower. In addition, away from closure, while capillarity imbibition gives a height proportional to the square root of time, gravity control imbibition leads to a linear relationship between the two magnitudes. Experimental data for the highly relevant silicon alloys/carbon system are discussed within the present framework.

\section{Modelling}
We shall consider   infiltration into a vertical capillary of cylindrical shape but with a radius that depends on height $z$ and time $t$,  $R(z,t)$, preserving axial symmetry. The fact that we use a system with axial symmetry does not preclude the extrapolation of the present analysis to the more general (and, actually, realistic) case of imbibition into porous preforms.
Capillarity, gravity and viscosity effects will be included. Temperature effects, that may in some cases be important \cite{MH98}, are neglected and contact angle assumed to be constant (independent, for instance, of infiltration speed, see Ref. \cite{Dg85,As02}).  
Assuming a Poiseuille-like dynamics and excluding kinetic effects in the equation of motion,  the   $z$-component of the fluid speed can be written as, 
\begin{equation}
v_z(r,t) = \frac{1}{4\mu}\frac{{\rm d}p}{{\rm d}z}\left[R^2 (z,t)-r^2\right]
\end{equation}
\noindent where $\mu$ is the liquid viscosity and $r$ is the radial variable at a horizontal section of the capillary. Mass flow can be written as,
\begin{equation}
Q(z,t)=2\pi\int_0^{R(z,t)}rv_z(r,t){\rm d}r = -\frac{\pi\left[R(z,t)\right]^4}{8\mu}\frac{{\rm d}p}{{\rm d}z}
\end{equation}
\noindent On the other hand,
\begin{equation}
Q=\pi  R_0^2{\dot z}_m
\end{equation}
Hereafter we shall use the following notation, $R(z_m,t_m)=R_m=R_0$, where $R_0$ is the capillary radius before infiltration has been initiated. 
\begin{figure}
\begin{center}
\includegraphics[width=7cm]{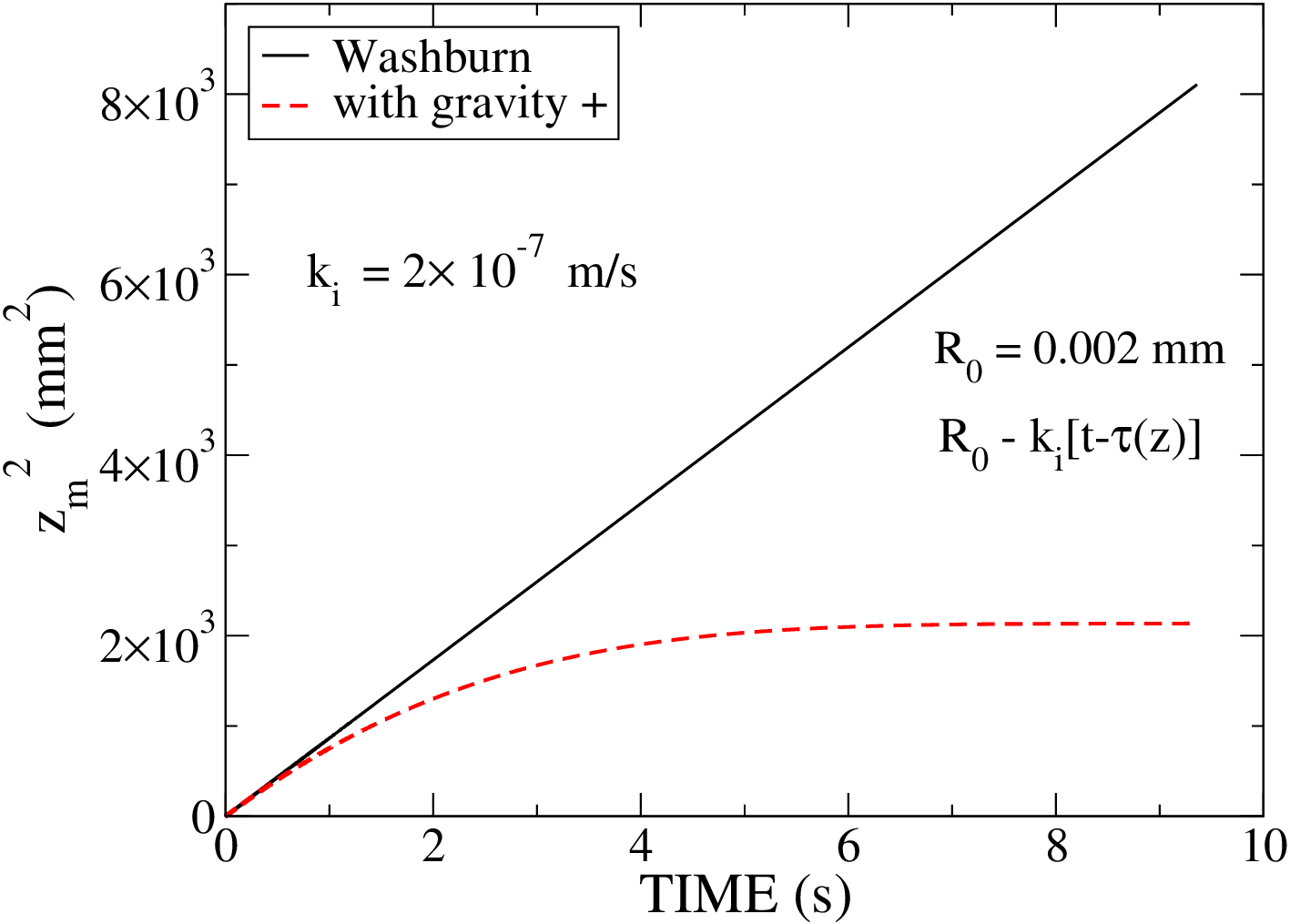} 
\vspace{1.5cm} \\
\includegraphics[width=7cm]{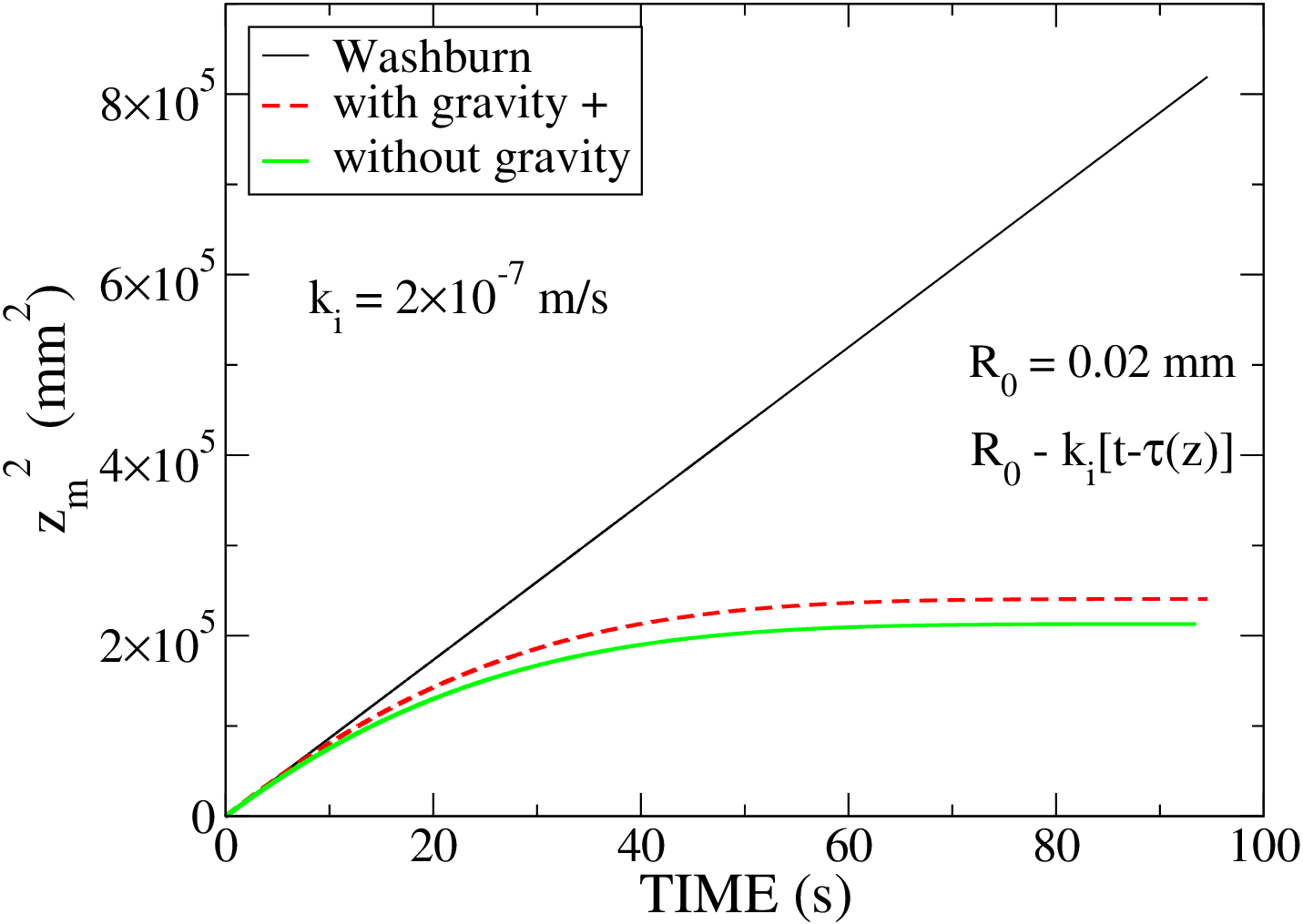}
\vspace{1.5cm} \\
\includegraphics[width=7cm]{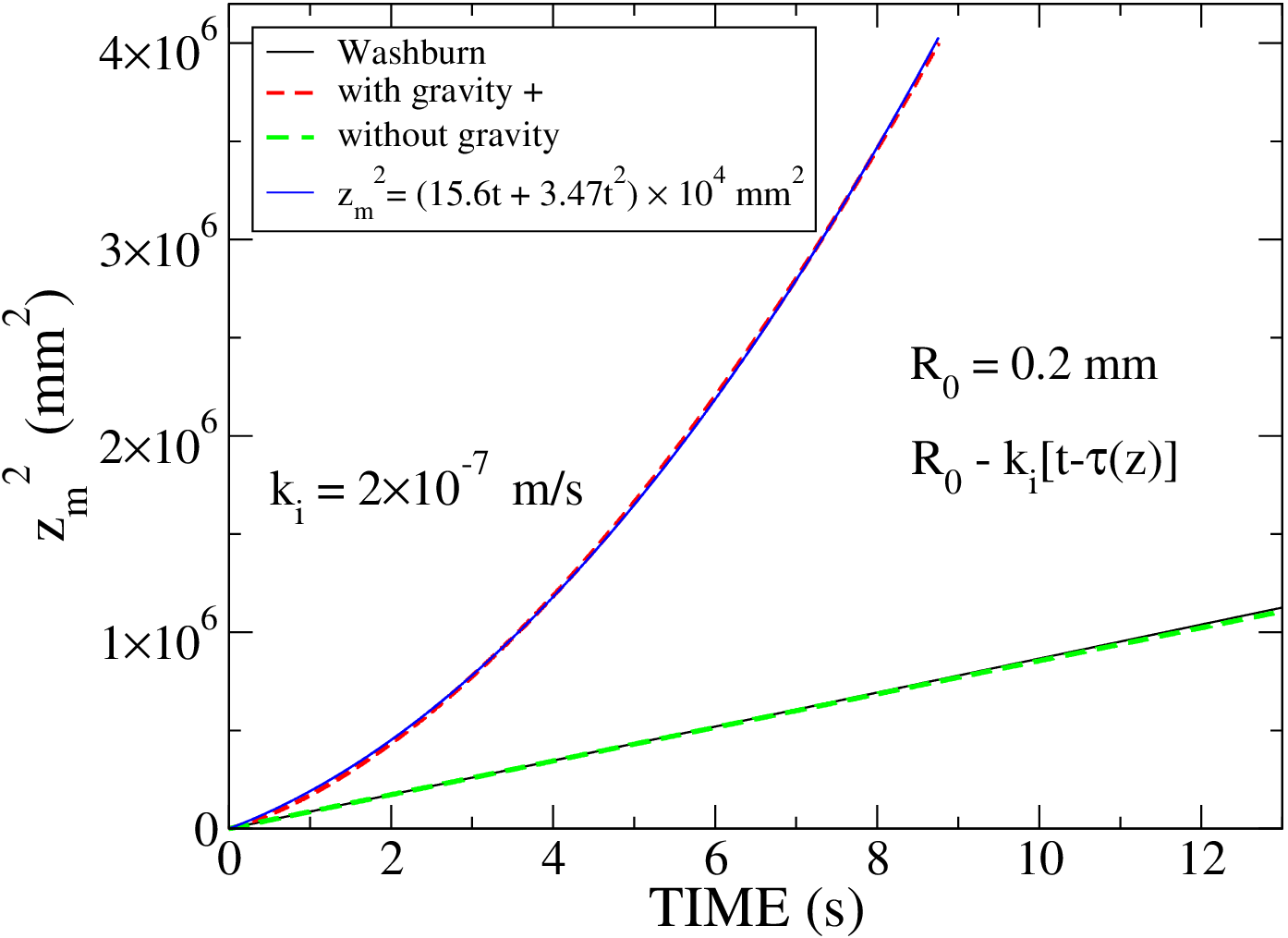}
\caption{Infiltrated height calculated in a capillary having a  radius initially of $R_0$=0.002 (upper) 0.02 (middle) and 0.2 (lower) mm,  assumed to diminish along the infiltration process with an interface-like law. The radius depends on both time $t$ and height $z$, see text. Results obtained with and without gravity are shown (in the case of $R_0$=0.002 mm both coincide).}
\label{S-p-vs-x}
\end{center}
\end{figure}
 Mass conservation imposes Equations (2) and (3) to be equal. Then, the pressure drop due to viscosity is,
\begin{equation}
p(z,t)=-8\mu R_0^2{\dot z}\int_0^{z}\left[R(z',t)\right]^{-4}{\rm d}z'
\end{equation}
\noindent This pressure drop must be compensated by the pressure at the liquid entrance $z$=0 and at the meniscus $z_m$ given by,
 \begin{equation}
p(0)=\mp \rho g z_m,\; \;\;\; p(z_m)=-\frac{2\gamma{\cos}\theta}{R_0}\;, 
 \end{equation}
 \noindent where $\rho$ is the density of the liquid and $\gamma$ its liquid-vapor surface tension. $\theta$  denotes the contact angle at the triple point. Then,
 \begin{equation}
\frac{2\gamma{\cos}\theta}{R_0} =8\mu R_0^2{\dot z_m}\int_0^{z_m}\left[R(z',t)\right]^{-4}{\rm d}z' \mp \rho g z_m
\end{equation}
\noindent and,
\begin{equation}
2z_m{\dot z_m} = \frac{1}{P}\left (B\pm z_mD\right)\;,
\end{equation}
  \noindent where the sign + (-) stands for downwards (upwards) infiltration and the constants $B$ and $D$ are given by,
  \begin{equation}
 B=\frac{R_0\gamma{\rm cos}\theta}{2\mu},  \;\;\;  D=\frac{R_0^2\rho g}{4\mu}\;\;,
 \end{equation} 
 \noindent where $P$ is an adimensional function of time given by,
 \begin{equation}
 P(t) = \frac{1}{z_m}\int_0^{z_m} \left[\frac{R_0}{R(z,t)}\right]^{4}{\rm d}z\;.
\end{equation}

In general, Eq. (8) has to be solved numerically assuming an explicit function for $R(z,t)$. A widely assumed function is,
 \begin{equation}
R(z,t)=R_0-k\left [t-\tau (z)\right ]^{\beta}\;,
 \end{equation}
 \noindent where $t$ is the actual time and $\tau (z)$ the time at which the liquid front passed through height $z$ so that $t-\tau (z)$ is the reaction time
 at $z$. This function becomes either the interface-like law ($\beta=1$ and $k=k_i$) or the diffusion-like law ($\beta=0.5$ and $k=k_d$), where $k_i$ and $k_d$ are constants usually quite different amongst them and varying significantly when either system (liquid/solid) and/or imbibition conditions. 
 
  \begin{figure}
\begin{center}
\includegraphics[width=7cm]{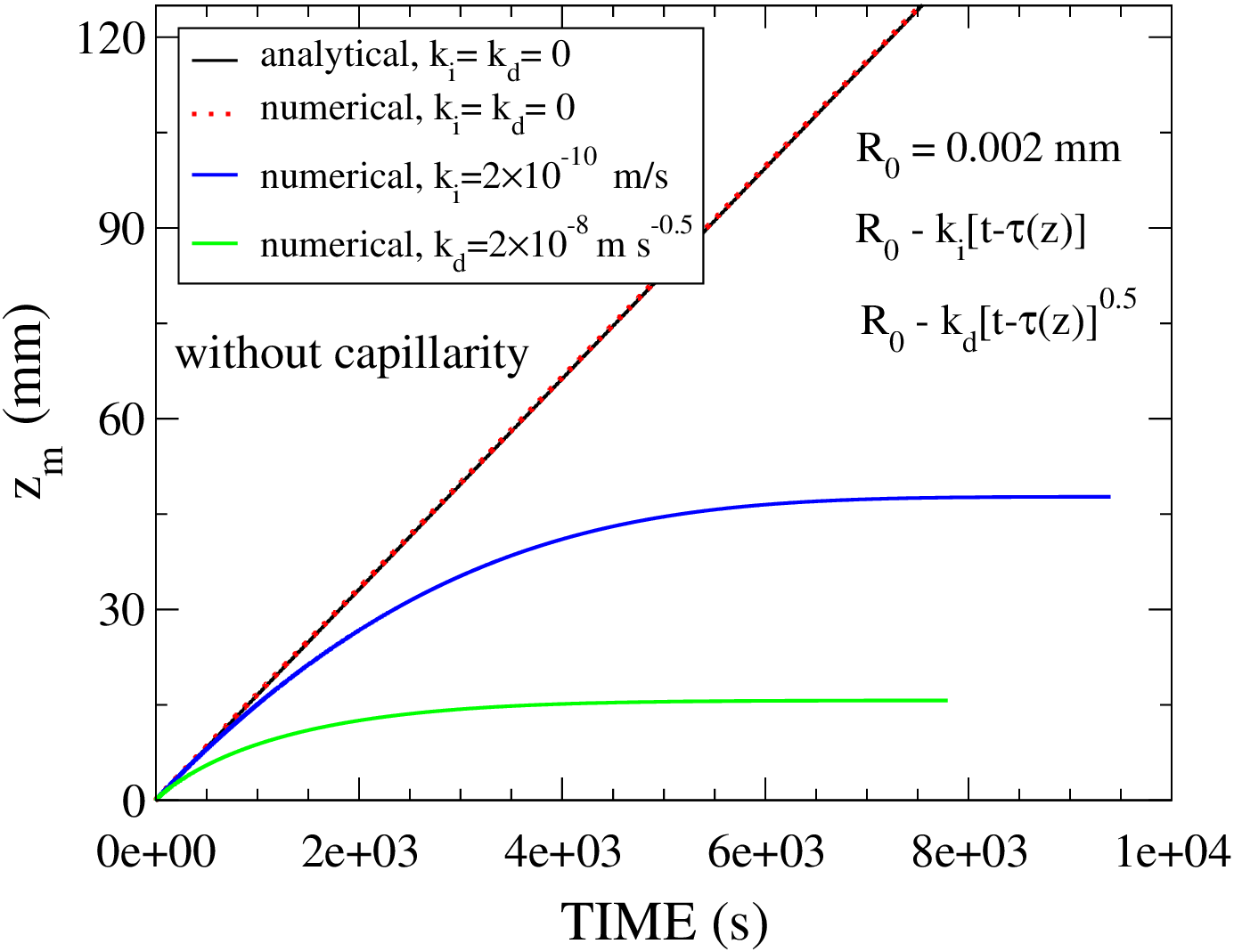} 
\caption{Calculated infiltrated height in a capillary having a  radius initially of $R_0$=0.002 mm, and assumed to diminish along the infiltration process with either an interface- or a diffusion-like law (reaction constants much smaller than those given in the text). Results correspond to imbibition dominated by gravity (a contact angle close to $90^\circ$). The radius depends on both time $t$ and height $z$.}
\label{S-p-vs-x}
\end{center}
\end{figure}

In most industrial or laboratory procedures  either the porous sample is immersed in the liquid or a liquid drop is deposited onto the solid. In both cases an additional  metallostatic pressure that may be non-negligible, shoud be included. Let us consider the case a drop having a height of $h$. Eq. (7) has to be rewritten as,
\begin{equation}
2z_m{\dot z_m} = \frac{1}{P}\left [B+ (h+z_m)D\right)]\;,
\end{equation}
\noindent Note that the new term acts as the capillarity term as it does not depend on $z_m$. Accordingly, the BCWL law has to be changed (see below).
 
\section{Results and Discussion}

  Hereafter we shall present and discuss resuts  for a canonical case: imbibition of liquid Si into a vertical carbon capillary. Parameters will be varied around  those usually accepted for this system. In particular the following values of the various constants in the preceding equations  were taken. The reaction constants in Eqs. (12)-(14)  $k_i=4\times10^{-8}$ m s$^{-1}$ and $k_d=2\times10^{-7}$ m s$^{-1/2}$ \cite{AS05}. The liquid-vapor surface energy for silicon melt $\gamma$ is known to vary in the range 0.72-0.76 N m$^{-1}$ \cite{Mi82} and its viscosity  $\mu$   0.46-0.76$\times10^{-3}$ Pa s \cite{TI74}; actually in this work we took $\gamma$=0.76 N/m and $\mu$=0.76 mPa s. The density of liquid Si near the melting point is 2.57 g cm$^{-3}$. The wetting angle at the Si/C interface in vacuum  has been reported to vary  from 0 to 22 deg \cite{WA75}, and, more recently \cite{BV06,VB08} to be 30 deg. In a rather recent review \cite{DE12} the contact angle for Si on graphite was shown to vary from near to $90^\circ$ at very short times, down to $30-15^\circ$ for contact times between liquid Si and graphite above 100 s. Actually, the latter value correspond to the case in which the Si/C interface has been replaced, due to reaction,  by  the Si/SiC interface. It is worth noting that  in capillaries with diameters larger than 0.01 mm \cite{SC16} gravity dominates imbibition and capillary closure does not occur for reasonable values of infiltrated  height and time. 

In this work we shall only consider the case in which imbibition occurs from above, that is gravity favors imbibition.  In the case that gravity works against capillarity (liquid infiltration from bottom upwards) the reasoning followed in the last subsection of the present section may need surely to be modified. Anyhow, data for infiltrated height versus time in that case is almost inexistent.   

\subsection{Imbibition controlled by capillarity}
As the results of Fig. 1 indicate, the  values of the  parameters given above lead to the case in which capillarity  dominate. In this case the limiting law is that referred to above as BCWL's law. As already discussed, it is obtained In the absence of gravity and reaction (null reaction constants and, thus,  $R(z,t)=R_0$ for all $z$ and $t$),  
 \begin{equation}
z_m^2 = (B+hD)t\;,
 \end{equation}
 \noindent where the effect of the drop height $h$ has been included (actual BCWL law is rescued for $h$=0). For the values of the system parameters given above, a capillary radius of 0.002 mm  and a drop height $h$=1.0 mm, the constants of this equation take the values, $B$ = 0.866 mm$^2$/s and $hD$= 0.033  mm$^2$/s.

Fig. 1 shows numerical results along with those given by BCWL's law  for  capillaries with radii 0.2, 0.02 and 0.002 mm. Albeit these sizes might seem too small, they are realistic when reactive infiltration of a variety of porous media is considered \cite{BV06,VB08}. Besides, varying the diameter in that range is sufficient to illustrate the possibilities of the approach proposed here. 
The results clearly  show the significant role that gravity plays. Numerical results with and without gravity coincide for the smallest capillary. On the  other hand, capillary closure occurs at times that are shorter as the capillary diameter decreases. In the largest capillary, however,  if gravity is neglected the numerical results closely follows BCLW's up to rather long imbibition times, capillary closure occurs beyond longer times (not shown in the Figure). Including gravity greatly increases imbibition performance reaching infiltration heights well above the classical BCLW law (lower panel in Fig. 1). Actually, the results for the square of the infiltrated height vesus time can be very accurately fitted by a 2nd order polynomium (see inset in the lower panel of Fig. 1). As the fitting indicates, at short times the BCWL's law holds, however, beyond 2 s the linear relation between infiltrated height and time, characteristic of gravity control imbibition (see below), increasingly dominates. Finally, it is worth noting that the main effects of replacing the interface reaction law by the diffusion law is to sharply increase infiltration height  delaying capillary closure.

\subsection{Gravity controlled imbibition}
Another limiting case of interest is that in which infiltration is control by gravity. Assuming a constant radius and taking $\theta=90^\circ$, we obtain,
 \begin{equation}
z_m = \frac{D}{2}t\;,
 \end{equation}
  \noindent The numerical results obtained by introducing a contact angle close to $90^\circ$ reproduce the linear relationship between the infiltrated height given by this equation,  provided that the capillary radius is kept constant.   In addition,  the imbibition rate is at  least two orders of magnitude higher than that obtained when imbibition is  controlled  by capillarity. Results for radii varying according to both  interface- and  diffusion-like law are shown. In the latter case the reaction constant was two order of magnitudes smaller than in the former, as decreasing $k_d$ in two orders of magnitude would give results  almost identical to those of Eq. (12) within he time range explored  here. This illustrates once more the expected delay in the growth of the reaction layer that the diffusion-like law introduces with respect to the interface-like law. It should be noted that gravity was introduced previously by several authors \cite{ST99} but without pointing out that when it controls imbibition, infiltrated length and time are linearly related.
 
 \begin{figure}[t!]
\begin{center}
\includegraphics[width=7cm]{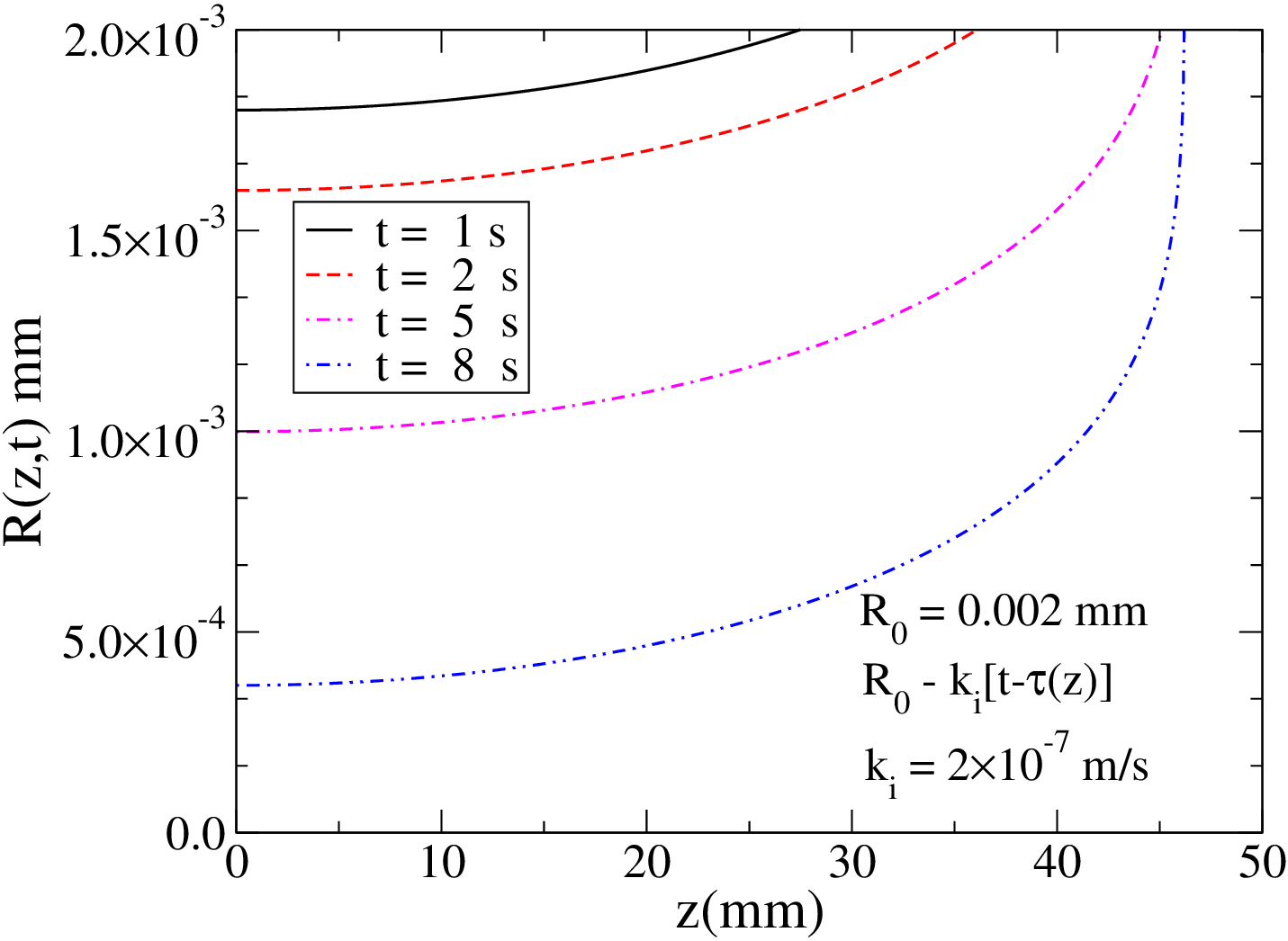}
\vspace{1.5cm} \\
\includegraphics[width=7cm]{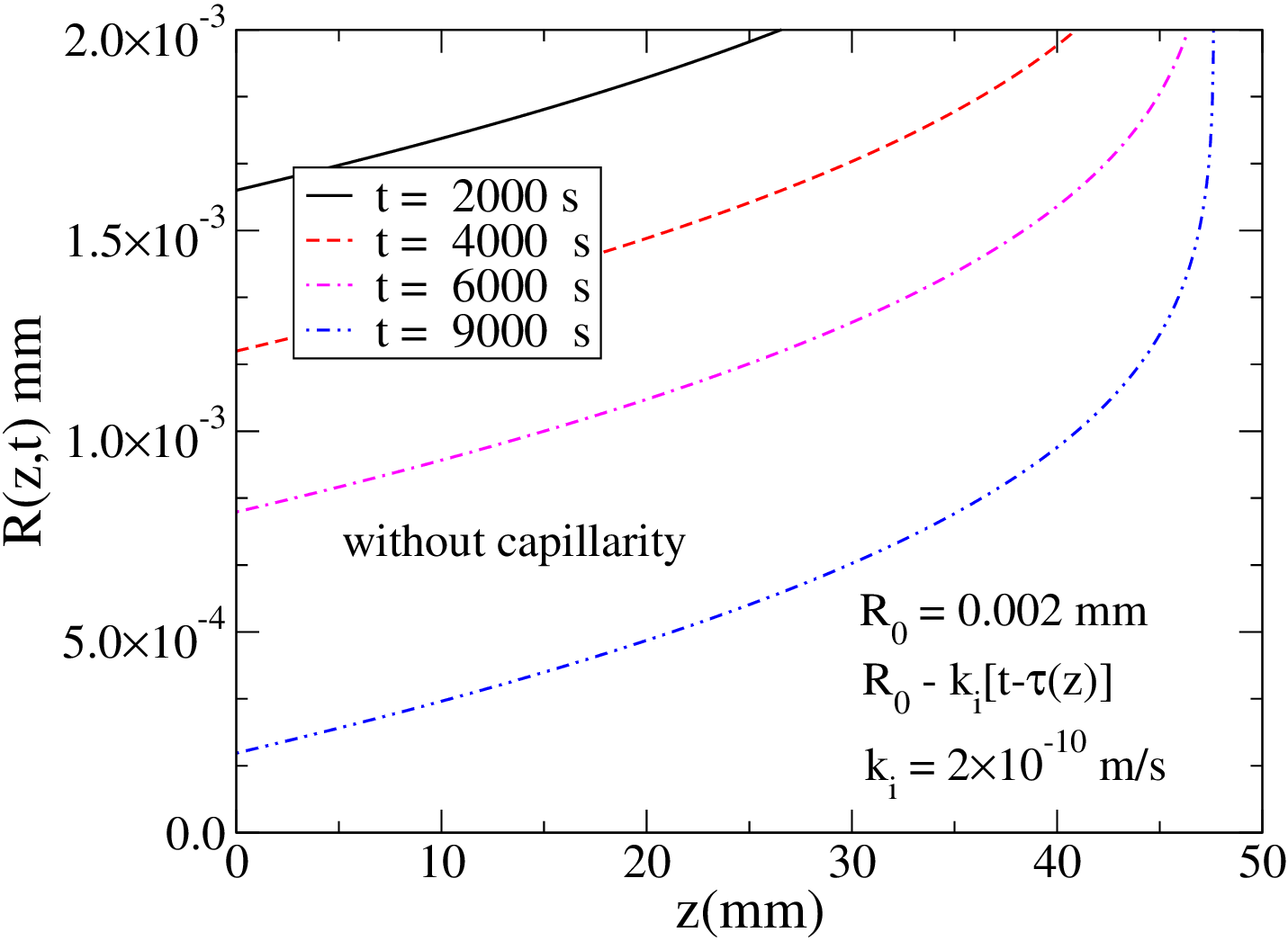}
\caption{Capillary radius versus infiltrated height for several infiltration times. The radius is assumed to vary  according to  an interface-like law. Results with  (upper panel)  and without (lower panel) capillarity are shown. Reaction constants are given in the Figure, whereas the rest of the parameters values are  those used in Figs. 1 (upper panel) and 3.}
\label{2Cylind}
\end{center}
\end{figure}

\subsection{Capillary Closure}

Fig. 3 shows results for the capillary radius as a function of height and four  infiltration times  for the capillary with initial radius of 0.002 mm that is assumed to be reduced as imbibition proceeds according to an  interface-like law. Results with (upper panel)  and without (lower panel) capillarity are shown. Reaction constants are given in the Figure, whereas the rest of the parameters values are  those used in Figs. 1 (upper panel) and 3. In both cases capillary closure occurs within the time ranges of Figs. 1 (upper panel) and 3. The results accounts  for an obvious feature: the thickness of the reaction layer increases (the capillary radius $R(z,t)$ decreases) with infiltration time. The dependence of the radius on the infiltrated height is clearly different in the two cases shown in the Figure. Actually, this strongly depends on the model parameters and no universal trend can be anticipated.
 
 \subsection{Can the present model contribute to the understanding of reactive infiltration?}
Let us start summarizing the present understanding of reactive infiltration in the canonical system Si/C. In a couple of recent  review papers Eustathopoulos et al \cite{EI10,Eu15} reviewed the literature on that system. From their analysis it may be concluded that:  i) the contact angle evolves from near $90^\circ$ just after Si was melted, down to 15 deg beyond approximately 300 s of contact time, ii) infiltration starts after an "incubation time" of 100 s, iii) an abrupt jump of 0.1 mm is followed by a steady linear infiltration from 0.1 up to 0.7 mm in approximately 250 s, iv) this gives a constant infiltration rate of 0.0025 mm/s. 

The linear relationship between infiltration length and time has been also reported to occur in reactive infiltration of NiSi alloys into porous graphite \cite{VB08}. Increasing the Si content from 47at\% up to 67at\% increased only in around 30\% the infiltration rate. However, an increase in temperature of only 100$^o$C (actually from 1170 up to 1270 $^o$C) raises infiltration rate in an order of magnitude (from 0.0001 mm/s up to 0.001 mm/s). In addition, beyond 3000 s at 1270 $^o$C, experimental data begin to deviate  from the linear law showing clear signs of saturation surely related to voids closure (see below and Ref. \cite{VB08}).

In trying to use the present framework to shed light into the experimental results for reactive systems it is first noted that there are very few experimental studies of density and viscosity of most alloys at high temperatures. This is particularly true for NiSi alloys \cite{AA12,LC12}. Anyhow, let us  make a zero-order approach to the reactive infiltration  problem aiming  to get at least a qualitative picture. From the above discussion, it is obvious that only when gravity controls imbibition a linear relationship between infiltrated length and time holds. Then, infiltration under reactive conditions may take place as follows: i)  In most cases, the contact angle is Initially larger than  $90^\circ$, ii) after some time, that may actually depend on the system under study and experimental conditions (initial contact angle and height of the drop on top of the porous preform), reaction gradually replaces the interface liquid/preform by the liquid/reaction product interface and, concomitantly, the contact angle  gets closer to $90^\circ$ (this may account for the "incubation time" or time elapsed before infiltration starts observed in many systems \cite{Eu15}), iii) beyond that time, gravity triggers infiltration  (actually  the contact angle does not need to be smaller than $90^\circ$ for the initiation of infiltration due to the metallostatic pressure exerted by the liquid metal on top of the porous preform, what is only required is that the constant in Eq. (2) vanish), iv) thereafter infiltration in controlled by gravity, provided that, due to reaction, the contact angle remains close to that at which infiltration was initiated (it is widely accepted that atoms of the liquid diffuse over the solid surface  ahead of the main infiltration front, favoring a decrease of the contact angle).

The results shown in Fig. 4 are a tentative approach to those depicted in Fig. 6 of Ref. \cite{VB08}. The latter Figure shows the infiltration height versus time of a liquid Ni47at\%Si alloy into a porous graphite preform of average pore radius around 2.5 microns, at temperatures in the range 1170-1270$^o$C. Some results are also presented for infiltration of Ni67at\%Si. In simulating this system we consider imbibition into a capillary of radius 0.4 microns (tortuosity may justify 
choosing a radius much smaller than the pores in the material used in the experiments) of a liquid of density 6.0 g cm$^{-3}$ (the possible temperature dependence of density was neglected) and a viscosity varying from 10-1 mPa s to account for the temperature variation just mentioned. This large decrease in viscosity upon an increase of temperature of only 100$^o$C may be unreasonable. However, the results of molecular dynamics for NiSi alloys reported in Ref. \cite{LC12} indicate that, in these alloys, viscosity may vary  faster with temperature than in pure materials. Finally, reaction was assumed to be interface controlled (at longer times it may be replaced by a diffusion-like law) with a rather small constant (see Fig. 4). The numerical results reproduce satisfactorily the experimental results depicted in Fig. 6 of Ref. \cite{VB08}. This agreement was achieved using parameters that may well be not too far from those unfortunately not yet measured.

\begin{figure}[t!]
\begin{center}
\includegraphics[width=7cm]{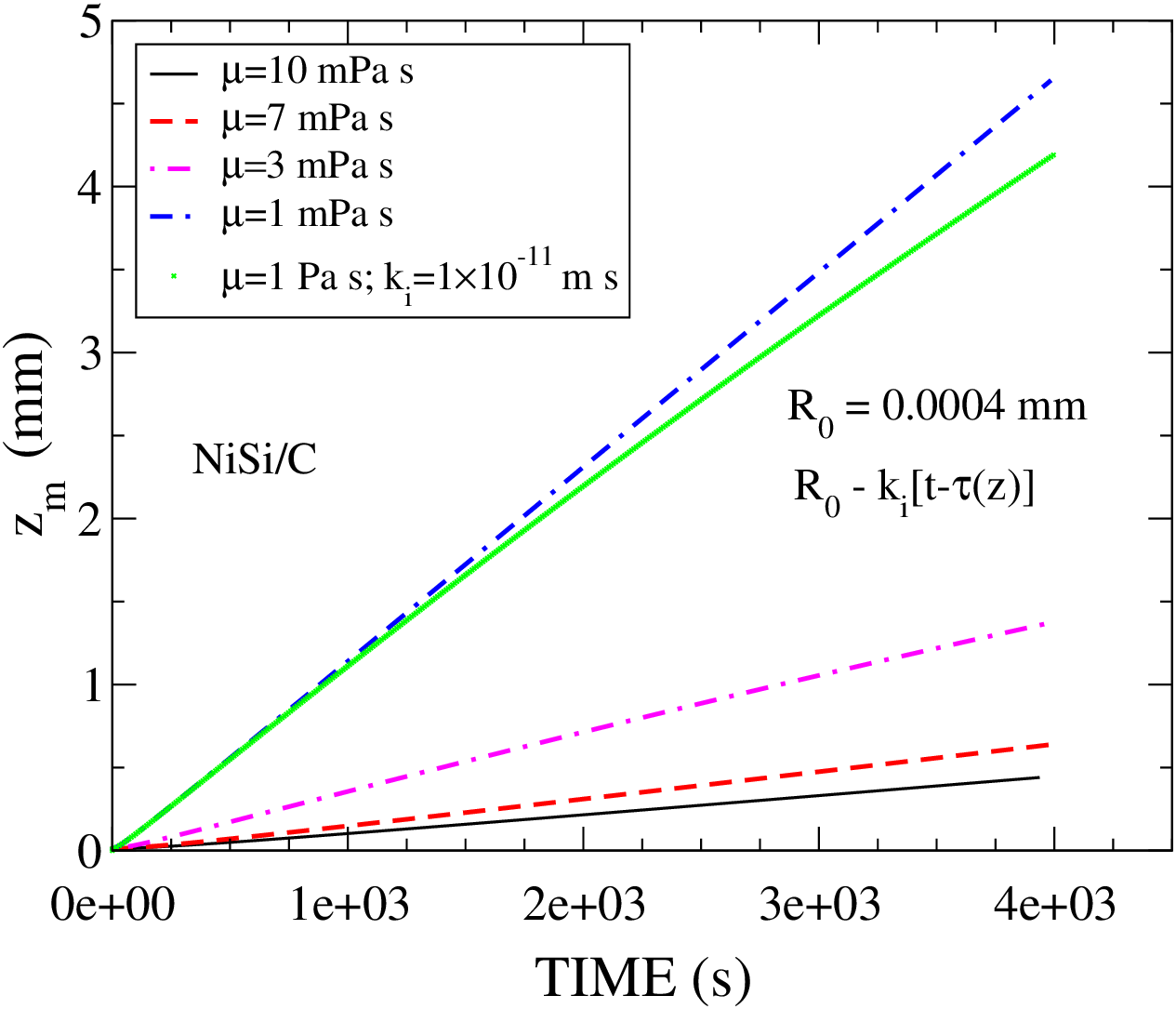}
\caption{infiltrated height versus infiltration time in a capillary of radius 0.0004 mm. Such small capillary may account for the "tortuosity" characteristic of most  porous carbon preforms used in industrial and research laboratories. Results for viscosities in the range 10-1 mPa s are depicted aiming to simulate the experiments of liquid NiSi infiltration into porous graphite at temperatures in the range 1170-1270$^o$C (see \cite{VB08}). In the case of the smallest viscosity the radius was assumed either to be  constant or vary  according to  an interface-like law. The same rate used for the remaining three cases does not introduce any change (note, however, that if time is increased considerably capillary closure will always occur).}
\label{2Cylind}
\end{center}
\end{figure}

Before ending is worth remarking on a serious flaw of the calculation reported above that concerns viscosity. In order to explain the temperature dependence of the infiltration  rate we assumed an increase in viscosity of an order of magnitude. This implies an activation energy for viscous flow similar to that found for reactive infiltration in \cite{VB08}, that is, around 230 kJ/mol. Experimental results for pure Si reported in \cite{ST95} were around 20 kJ/mol, in fair agreement with the molecular dynamics result of 16 kJ/mol \cite{LC12}. The authors of the latter  work obtained 24 kJ/mol for pure Ni. Both are at least an order of magnitude below the just mentioned value for reactive infiltration. What about metallic alloys?. Although the viscosity of pure metals is still a subject of research, is just a matter of improving precision. This is not the case of metallic alloys and, very specially, those undergoing a glass transition \cite{LC12,BE15,DH16}. Ni$_x$Si$_{100-x}$ alloys belong to this group for particular values of $x$.  Actually, those studied experimentally and by means of molecular dynamics simulations ($x$=20 and 25, see Refs.  \cite{LC12,DH16}), show that transition.  Molecular dynamics simulations predicted a change in dynamics from an Arrhenius-type behavior to a power-law, for  temperatures above the glass transition temperature \cite{LC12,DH16}, in agreement with experiments for those alloys \cite{DH16}. Such a crossover has been also reported to occur in a wide range of binary, ternary and quaternary alloys,  all having in common a glass transition \cite{BE15}. Using the theoretical data  reported in \cite{LC12} we estimated an activation energy of 45 kJ/mol for the two alloys investigated in that work, still a factor of 6 smaller than the activation energy for reactive infiltration \cite{VB08}. Although no experimental data for the alloys considered here ($x$=53 and 33) are available, they will also presumably undergo a glass transition. In such a case, experimental data are required before our ansatz (a large activation energy) can be discarded.

 \section{Concluding Remarks}
The present approach provides a reasonable mechanism for reactive infiltration, a fabrication method whose functioning has worried  many scientists in the field form any years. The first step was  realising that only gravity controlled imbibition leads to: i) a linear relationship between infiltrated length and time, and, ii) a rather low infiltration rate. Capillarity controlled infiltration fullfils neither of those two features. Once, this was solidly settled, reasons supporting the possibility that reactive infiltration might be controlled by gravity had to be found. Having noted that in these systems the initial contact 
angle is greater than $90^\circ$ (non-wetting condition) and that decreases with contact time due to reaction  towards  wetting, we realised that before the contact angle gets smaller than $90^\circ$, the capillarity force may be compensated by the metallostatic  force due to the liquid drop on top of the porous preform, leaving gravity as the only agent to initiate infiltration.   

Thereafter infiltration is controlled by gravity, as, due to the fact that upon infiltration the metal should get in touch with a slightly reacted surface (some atoms of the liquid may diffuse ahead of the infiltration front)  the contact angle may likely remain close to that at which infiltration was initiated. However, such a simplification left viscosity as the only parameter capable of reproducing the activation energy of reactive infiltration. Consequently we assumed that the viscosity of the NiSi alloys considered here decreased an order of magnitude over the temperature range of 1170-1270$^\circ$C. This seems a huge change if one recalls what is known for liquid pure metals and alloys. In order to justify our assumption, we have argued  that it is likely that the two NiSi undergo a glass transition and, consequently, show some important anomalies.
The fact that, up to our knowledge, there  is no alternative approach explaining the linear relationship between infiltrated length and time, 
 justifies experimental studies addressed to determine the activation energies for viscous flow in these alloys.
\vspace{5mm} 

\noindent{\bf Acknowledgments}
\vspace{5mm} \\
This work has been partially supported by the Spanish Ministerio de Econom{\'i}a Industria y Competitividad  (MAT2016-77742-C2-2-P and AYA2015-66899-C2-2-P). 


\end{document}